\documentclass{ws-procs9x6}            
\begin{document}
\title{The role of $X(4140)$ and $X(4160)$ in the reactions of $B^+ \to J/\psi \phi K^+$}

\author{En Wang$^*$}
\address{School of Physics and Microelectronics, Zhengzhou University, Zhengzhou, Henan 450001, China\\
$^*$E-mail: wangen@zzu.edu.cn}

\author{Ju-Jun Xie}
\address{Institute of Modern Physics, Chinese Academy of
Sciences, Lanzhou 730000, China, and\\
School of Physics and Microelectronics, Zhengzhou University, Zhengzhou, Henan 450001, China}

\author{Li-Sheng Geng}
\address{School of Physics, Beihang University, Beijing 100191, China, and\\
School of Physics and Microelectronics, Zhengzhou University, Zhengzhou, Henan 450001, China}

\author{Eulogio Oset}
\address{Departamento de
F\'{\i}sica Te\'orica and IFIC, Centro Mixto Universidad de
Valencia-CSIC Institutos de Investigaci\'on de Paterna, Aptdo.
22085, 46071 Valencia, Spain}

\begin{abstract}
We have studied the $J/\psi\phi$ mass distribution of the process $B^+\to J/\psi\phi K^+$ from the threshold to about 4250~MeV, by considering the contribution of the $X(4140)$ with a narrow width, together with the $X(4160)$ state. Our results show that the cusp structure at the $D^*_s\bar{D}^*_s$ threshold is tied to the molecular nature of the $X(4160)$ state.
\end{abstract}


\bodymatter

\section{Introduction}\label{sec:introduction}
Recently, the state $X(4140)$ was confirmed in the measurement of the $B^+ \to J/\psi \phi K^+$ reaction at LHCb~\cite{Aaij:2016nsc,Aaij:2016iza}. However, one surprise is that the $X(4140)$ deduced from the analysis, with quantum numbers $J^{PC}=1^{++}$, has a width $\Gamma \approx 83\pm 21^{+21}_{-14}$~MeV, substantially larger than that claimed in the former experiments~\cite{PDG2016}.

In this work, we show that the low invariant $J/\psi \phi$ mass region,  requires the contributions of a narrow $X(4140)$, and a wide $X(4160)$ resonance which couples to $J/\psi \phi$ but is mostly made by a $D_s^* \bar{D}_s^*$ molecule as discussed in Ref.~\citenum{Molina:2009ct}. As a consequence of analyticity and driven by the molecular nature of $X(4160)$, the $J/\psi\phi$ mass spectrum develops the strong cusp around the $D_s^* \bar{D}_s^*$ threshold~\cite{Aaij:2016nsc,Aaij:2016iza}.


\section{The formalism}
\label{sec: formalism}
\begin{figure}
\includegraphics[width=0.48\textwidth]{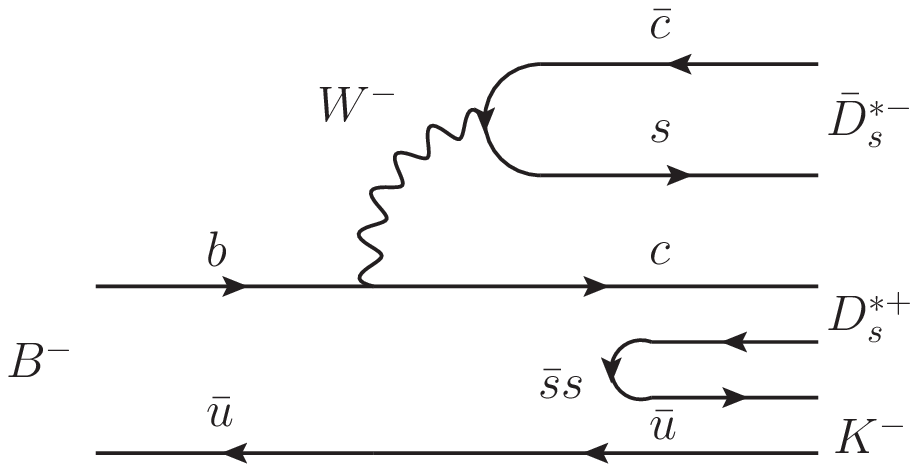}
\includegraphics[width=0.48\textwidth]{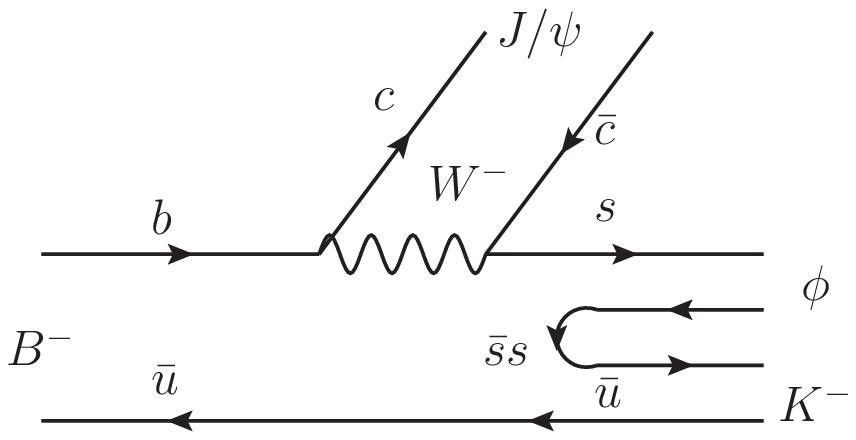}
\caption{Microscopic quark level production of $K^- D_s^*\bar{D}_s^*$ (left) and $B^- \to K^- J/\psi \phi$ (right) in $B^-$ decay.}  \label{fig:weakdecay}
\end{figure}
\begin{figure}
\includegraphics[width=0.6\textwidth]{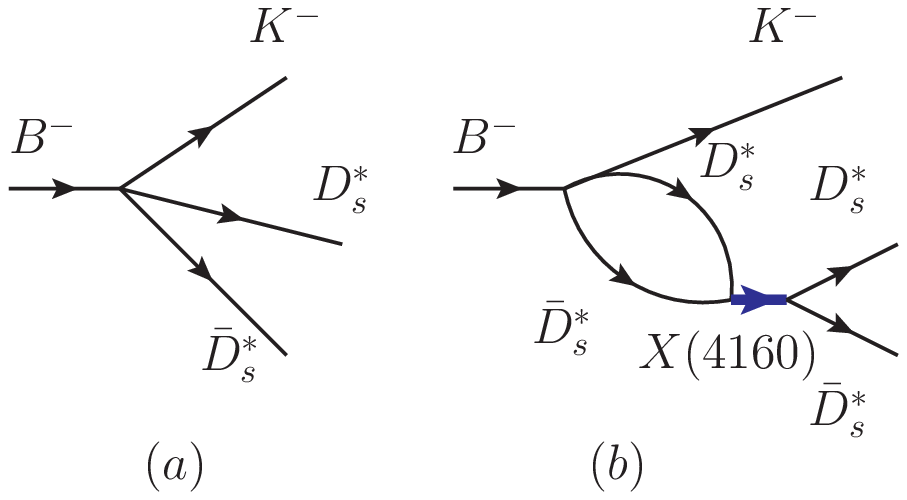}
\includegraphics[width=0.35\textwidth]{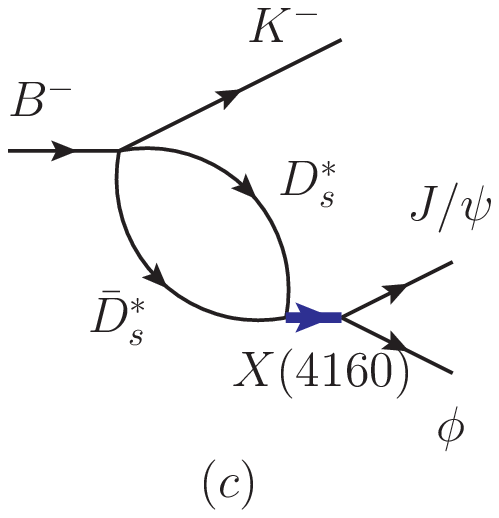}
\caption{(a) The tree diagram, and (b) the $D_s^*\bar{D}_s^*$ final state interaction for $B^- \to K^- D_s^*\bar{D}_s^*$ in the presence of the $X(4160)$ resonance. (c) The mechanism for $B^- \to K^- J/\psi \phi$ driven by the $X(4160)$ resonance.}  \label{fig:DsDs}
\end{figure}

As the first step, the dominant process of $B^- \to K^- D_s^*\bar{D}_s^*$ (we take the complex conjugate reaction to deal with $b$ quark rather than $\bar{b}$ quark) at the quark level proceeds as shown in the left panel of Fig.~\ref{fig:weakdecay}, involving external emission. Obviously in the neighborhood of the resonance the tree level term of Fig.~\ref{fig:DsDs}(a) is small compared to the resonant term of Fig.~\ref{fig:DsDs}(b), but we keep it in the calculations. For the production of $J/\psi \phi$ with this mechanism, the tree level of Fig.~\ref{fig:DsDs}(a) does not contribute and then we are led to the diagram of Fig.~\ref{fig:DsDs}(c)

The $B^- \to K^- J/\psi \phi$ reaction can also proceed through the mechanism as shown in the right panel of Fig.~\ref{fig:weakdecay}
involving internal conversion. Yet, the internal conversion is penalized by color factors with respect to the external emission of the left diagram of Fig.~\ref{fig:weakdecay}, and hence this term, or rescattering of this term like that in Fig.~\ref{fig:DsDs}(c), but with $J/\psi \phi$ intermediate state instead of
 $D_s^*\bar{D}_s^*$, which would involve the extra factor $g_{J/\psi\phi}/g_{D_s^*\bar{D}_s^*}$ versus the amplitude of \ref{fig:DsDs}(c), can be safely neglected.

We can then write the amplitude for the $B^-\to K^-D_s^*\bar{D}_s^*$,
\begin{eqnarray}
\mathcal{M}_{B\to K D_s^*\bar{D}_s^*} &=& A\left[ 1+G_{D_s^*\bar{D}_s^*}(M_{\rm inv}(D_s^*\bar{D}_s^*)) \right. \nonumber \\
&&\times \left. t_{D_s^*\bar{D}_s^*\to D_s^*\bar{D}_s^*}(M_{\rm inv}(D_s^*\bar{D}_s^*))\right] \left( \vec{\epsilon}\cdot \vec{k}\, \vec{\epsilon}^{\,\prime}\cdot \vec{k}-\frac{1}{3}\vec{k}^2\vec{\epsilon}\cdot \vec{\epsilon}^{\,\prime}\right), \label{eq:amp1}
\end{eqnarray}
and  the amplitude for the $B^-\to K^-J/\psi\phi$,
\begin{eqnarray}
\mathcal{M}_{B\to K J/\psi\phi}
 &=& A\times G_{D_s^*\bar{D}_s^*}(M_{\rm inv}(J/\psi\phi))\nonumber \\
&& \times  t_{D_s^*\bar{D}_s^*\to J/\psi\phi}(M_{\rm inv}(J/\psi\phi)) \left( \vec{\epsilon}\cdot \vec{k}\, \vec{\epsilon}^{\,\prime}\cdot \vec{k}-\frac{1}{3}\vec{k}^2\vec{\epsilon}\cdot \vec{\epsilon}^{\,\prime}\right) \nonumber \\
&& + \frac{B \, M^3_{X(4140)}(\vec{\epsilon}_{J/\psi} \times \vec{\epsilon}_\phi) \cdot \vec{k}}{M^2_{\rm inv}(J/\psi\phi)-M^2_{X(4140)}+iM_{X(4140)}\Gamma_{X(4140)}} \nonumber  \\
 &=& \mathcal{M}^{X(4160)} +\mathcal{M}^{X(4140)}\label{eq:amp2}
\end{eqnarray}
where $\vec{\epsilon}$, $\vec{\epsilon}^{\,\prime}$ are the polarization vectors of $D_s^*$ and $\vec{D}_s^*$ (or $J/\psi$ and $\phi$), and we evaluate it in the frame of reference where the $D_s^*\bar{D}_s^*$ (or $J/\psi\phi$) system is at rest. For the $J/\psi\phi$ production, we have included the term $\mathcal{M}^{X(4140)}$ to account for the production of $J/\psi\phi$ via the $1^{++}$ $X(4140)$ resonance. Here we take $M_{X(4140)}=4132$~MeV and $\Gamma_{X(4140)}=19$~MeV. The $A$ and $B$ are free parameters to be fitted to the data. The details of the loop functions $G$ and the transition amplitude $t$ in Eqs.~(\ref{eq:amp1}, \ref{eq:amp2}) are given in Ref.~\citenum{Wang:2017mrt}. Finally, the mass distribution can be written as,
\begin{eqnarray}
\frac{d\Gamma}{dM_{\rm inv}} &=& \frac{1}{(2\pi)^3}\frac{1}{4M_{B^-}^2}|\vec{k}'|\, \tilde{p}|\mathcal{M}|^2, \label{eq:dwidth}
\end{eqnarray}
where $\vec{k}'$ is the $K^-$ momentum in the $B^-$ rest frame, and $\tilde{p}$ the $D_s^*$ (or $J/\psi$) momentum in the $D_s^*\bar{D}_s^*$ (or $J/\psi\phi$) rest frame.

\section{Results}
\begin{figure}
\includegraphics[width=0.48\textwidth]{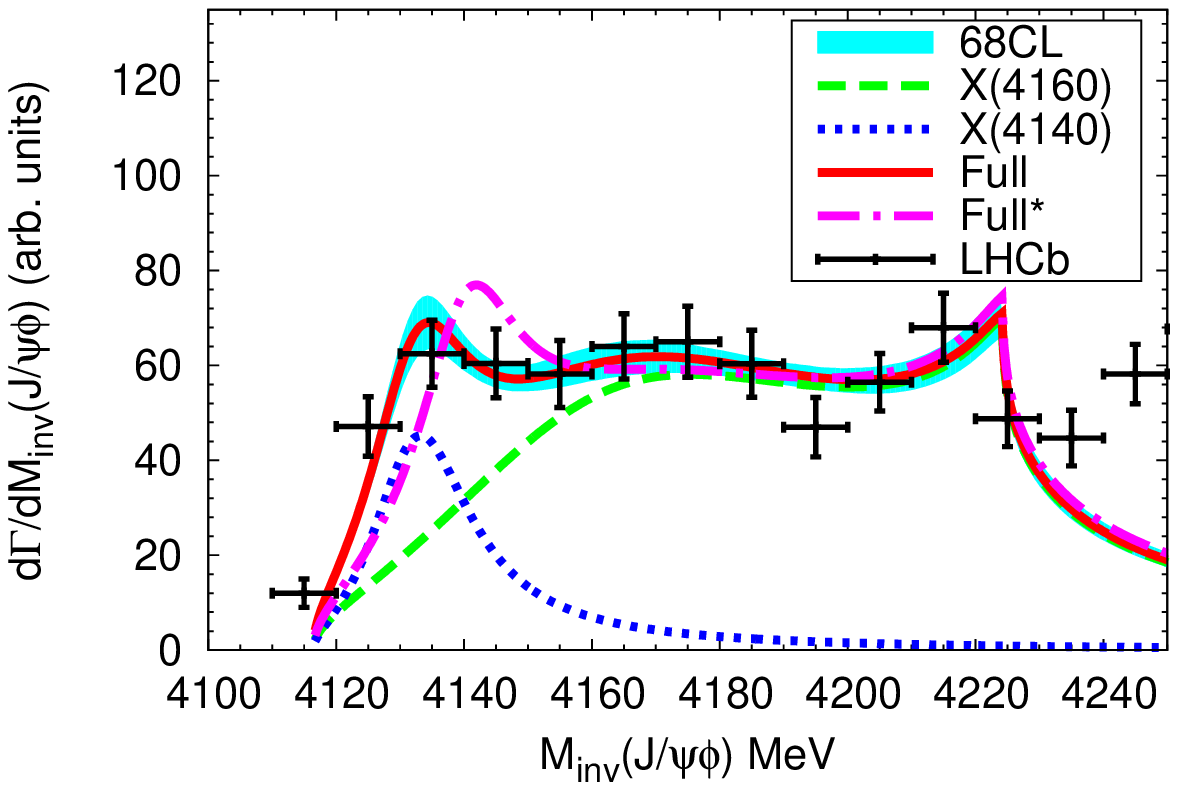}
\includegraphics[width=0.48\textwidth]{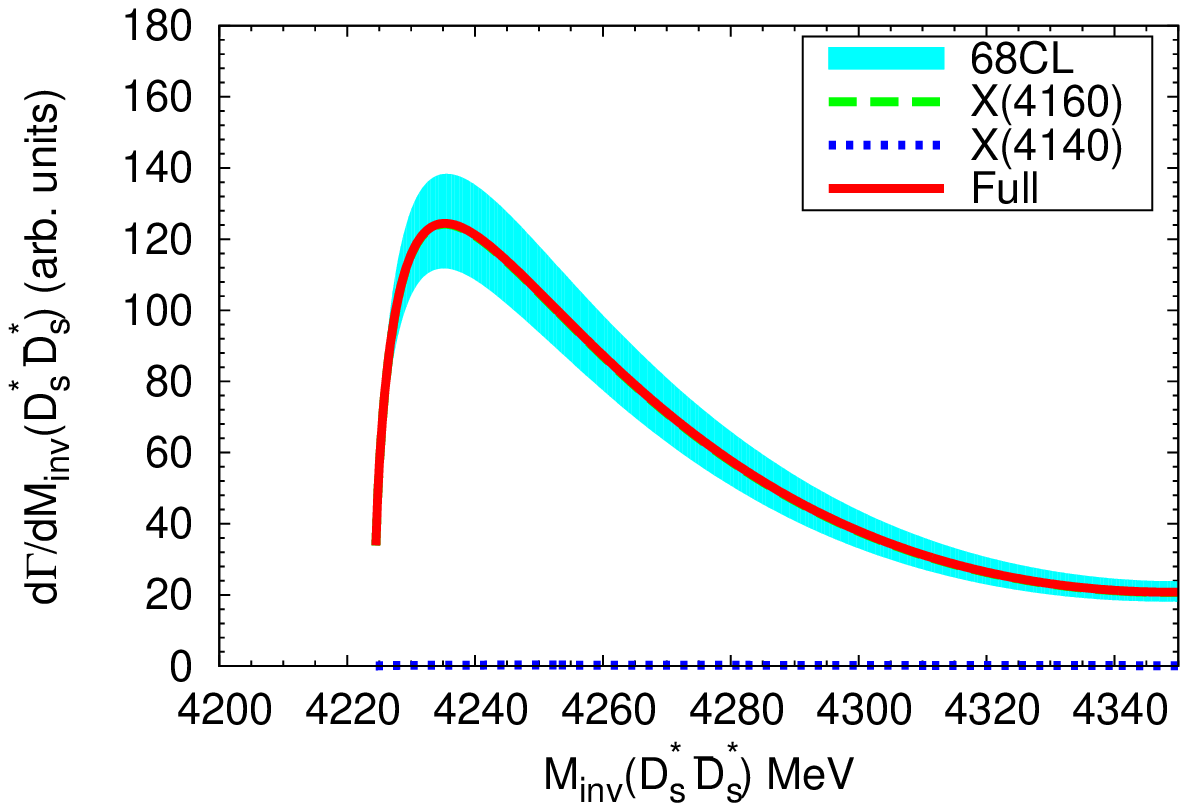}
\caption{Left: the $J/\psi\phi$ mass distribution of the  $B^- \to K^- J/\psi \phi$ decay. Right: The $D_s^*\bar{D}_s^*$ mass distribution of the $B^- \to K^- D_s^*\bar{D}_s^*$ decay. The bands come considering the errors in the fitted parameters, and represents the 68\% confidence-level.}
\label{fig:result_Jpsiphi}
\end{figure}

A suitable fit to the data is obtained as shown in the left panel of Fig.~\ref{fig:result_Jpsiphi}, Our results (red solid curve, labeled as `Full') are in good agreement with the experimental data at low $J/\psi\phi$ invariant masses, and we also present the uncertainties of our results in the figure.
As we can see in the figure, we obtain a contribution from the $X(4140)$ (blue dotted curve) that is dominant at low invariant masses, and is responsible for the peak observed in the experiment around 4135~MeV. The $X(4160)$  (green dashed curve) is responsible for most of the strength and produces a broader peak around 4170~MeV. And finally a cusp appears at the $D_s^*\bar{D}_s^*$ threshold as it shows up in the experiment. This cusp comes from the factor $G_{D_s^*\bar{D}_s^*}(M_{\rm inv})$ and reflects the analytical structure of this function with a discontinuity of the derivative at threshold. A similar feature is also found in Refs.~ \citenum{Dai:2018nmw,Wang:2018djr}. One must stress that this factor appears here as a consequence of analyticity, with a large strength due to the $D_s^*\bar{D}_s^*$ molecular structure of the $X(4160)$. In an analysis like the one of Refs.~\citenum{Aaij:2016nsc,Aaij:2016iza}, where a sum of amplitudes for resonance excitation and some background are fitted to the data, this factor is not considered, and as a consequence the cusp around $D_s^*\bar{D}_s^*$ in the data is missed in the fit.

To finish the work, and as a test of the explanation given here, we present in the right panel of Fig.~\ref{fig:result_Jpsiphi} the $D_s^*\bar{D}_s^*$ mass distribution above threshold obtained with the same parameters as in Fig.~\ref{fig:result_Jpsiphi}.
As we can see in this figure, there is a peak close to threshold, which should not be mis-identified with a new resonance, but it is the reflection of the $X(4160)$ which in our fit has the mass at 4169~MeV.

\section{Summary}
The present work and the prediction, tied to the interpretation given for the $B^-\to K^-J/\psi\phi$ spectrum, should act as an incentive to measure this reaction and learn about properties of the $X(4140)$ and $X(4160)$. In addition, the $J/\psi \phi$ mass distribution of  the $e^+e^- \to \gamma J/\psi\phi$ reaction reported by BESIII~\cite{Ablikim:2017cbv} is also compatible with the existence of the $X(4140)$ state, appearing as a peak, and a strong cusp structure at the  $D_s^*\bar{D}_s^*$ threshold, resulting from the molecular nature of the $X(4160)$ state\cite{Wang:2018djr}.

\section*{Acknowledgements}
This work is partly supported by the National Natural
Science Foundation of China under Grant Nos.  11375024, 11522539, 11735003, 11505158, 11475015, and  11647601.  It is also supported by
the Youth Innovation Promotion Association CAS (No.
2016367),
and Academic Improvement Project of Zhengzhou University.
This work is also partly supported by the Spanish Ministerio de Economia
y Competitividad
and European FEDER funds under the contract number FIS2014-57026-REDT,
FIS2014-51948-C2-1-P, and FIS2014-51948-C2-2-P, and the Generalitat
Valenciana in the program Prometeo II-2014/068.

\bibliographystyle{ws-procs9x6} 
\bibliography{ws-pro-sample}



\end{document}